\documentstyle[prc,aps,preprint,tighten,psfig]{revtex}
\begin{document}
\draft
\title{Three-potential formalism for the three-body Coulomb 
scattering problem \thanks{to be published in Phys. Rev. C {\bf
55}, March (1997)}}
\author{Z. Papp}
\address{Institute of Nuclear Research of the 
Hungarian Academy of Sciences, \\
P.O. Box 51, H--4001 Debrecen, Hungary}
\date{8 October 1996}
\maketitle

\begin{abstract}
\noindent

We propose a three-potential formalism for the three-body 
Coulomb scattering problem. 
The corresponding integral equations are 
mathematically well-behaved 
and can succesfully be solved by the Coulomb-Sturmian separable
expansion method. The results show perfect agreements with existing
low-energy $n-d$ and $p-d$ scattering calculations.
\end{abstract}

\vspace{0.5cm} 

\narrowtext

\section{INTRODUCTION}

Since the Faddeev equations are the fundamental equations 
of the three-body
problems their solutions are of central interest in many branches of
physics. This is especially true in nuclear physics 
because three-body
calculations serve as a distinguished tool for studying 
the fundamental
nucleon-nucleon interactions. A general interaction may 
have a local or
non-local short-range part and a long-range Coulomb part. 
The solution of
the Faddeev equations with such type of potentials is 
not an easy job,
especially the Coulomb interaction and the scattering 
dynamics make the
procedure very complicated. There exits extensive 
literature on the subject
(see, e.g., Refs. \cite{alt-naning,chandler,bencze1} 
and references therein)
so we restrict ourselves only to practical approaches.

There are two genuinely different approaches 
in the practical Faddeev
calculations that in some extent can handle 
Coulomb-like interactions in
scattering-state problems. One of them is based on 
the solution of the
configuration-space differential equations using the 
asymptotic boundary
conditions \cite{grenoble}. In the other approach, in order that the
standard techniques could be applied, the long-range 
Coulomb potential is
screened, and then, as the screened Coulomb potential 
goes to the unscreened
one, a renormalization procedure is applied \cite{alt-sandhas}. 
However, in
spite of the rapid development we have witnessed in the 
past few years, only
limited solutions below or above the breakup threshold 
are available yet
(see, e.g. Refs. \cite{chen,berthold,bzank}.

An another approach to the nuclear three-body problem with Coulomb
interaction were derived along the two-potential formalism. 
The first, and
formally exact, approach was proposed by Noble \cite{noble}. In this
formulation all the Coulomb interactions were 
included in "free" Green's
operator. Thus the corresponding Faddeev-Noble 
equations are mathematically
well-behaved and in the absence of Coulomb interaction 
they fall back to
the standard equations. However, the associated 
Green's operator is not
known, so this formalism is not suitable for 
practical calculations.

The aim of this paper is to treat the three-body Coulomb 
scattering problem
via the solution of the Faddeev-Noble integral equations. 
In Sec. II below we
shall derive a ''three-potential'' formalism. 
We will arrive at a set of
Lippmann-Schwinger and Faddeev equations which form 
a mathematically
well-behaved set of integral equations. 
In Sec. III below we shall describe
how the solution can be reached. 
In Sec. IV below we shall compare our
calculations with existing benchmark $n-d$ and $p-d$ below-breakup
scattering results. It is found that in all cases 
excellent agreement is
achieved. The method therefore appear as a promising and 
efficient tool for
solving the Coulomb three-body scattering problem, as it 
can be adapted to
more general cases and can be extended to above-breakup 
Coulomb scattering
calculations.

\section{THREE-POTENTIAL FORMALISM FOR THE THREE-BODY 
COULOMB SCATTERING PROBLEM}

The Noble's approach, which is, in fact, a two-potential 
formalism, requires
the knowledge of the complete solution of the three-body 
Coulomb problem.
Bencze has suggested to replace the incalculable 
three-body Coulomb Green's
operator by the channel-distorted Coulomb Green's 
operator \cite{benczenp}.
In this section below we will follow the derivation
of Ref.~\cite{benczenp}, but instead of neglecting 
the intermediate range
polarization potential, we will link the three-body Coulomb
Green's operator to the channel-distorted Coulomb 
Green's operator via a
Lippmann--Schwinger equation. Thus we will arrive 
at a set of Faddev-Noble
and Lippmann-Schwinger integral equations which 
are mathematically well-behaved 
because all the long-range interactions are kept in the Green's
operator.

The Hamiltonian of a three-body system with short-range plus Coulomb
two-body interactions reads 
\begin{equation}
H=H^0 + v_\alpha^s + v_\beta^s + v_\gamma^s + 
v_\alpha^C + v_\beta^C +
v_\gamma^C,
\end{equation}
where $H^0$ is the three-body kinetic energy 
operator, $v_\alpha$ denotes the
interaction in subsystem $\alpha$ and the 
superscript $s$ and $C$ stand for
short-range and Coulomb, respectively. 
We introduce here the usual
configuration-space Jacobi coordinates
 $\xi_\alpha$ and $\eta_\alpha$; $\xi_\alpha$ is the coordinate
between the pair $(\beta,\gamma)$ and $\eta_\alpha$ is the
coordinate between the particle $\alpha$ and the center of mass
of the pair $(\beta,\gamma)$.
Thus the potential $v_\alpha$, the interaction between the
pair $(\beta,\gamma)$, appears as $v_\alpha (\xi_\alpha)$.

The asymptotic Hamiltonian is defined as 
\begin{equation}
H_\alpha=H^0 + v_\alpha^s + v_\alpha^C,
\end{equation}
and the asymptotic states are the eigenstates of $H_\alpha$ 
\begin{equation}
H_\alpha | \Phi_{\alpha} \rangle = E | \Phi_{\alpha} \rangle,
\end{equation}
where $\langle \xi_\alpha \eta_\alpha |
 \Phi_{\alpha} \rangle = \langle
\eta_\alpha| \chi_{\alpha} \rangle \langle 
\xi_\alpha | \phi_{\alpha} \rangle
$, a product of a free motion in coordinate
 $\eta_\alpha$ and a bound-state
in the two-body subsystem $\xi_\alpha$.

We define two asymptotic Coulomb Hamiltonians as 
\begin{equation}
H_\alpha^C=H^0 + v_\alpha^s + v_\alpha^C + v_\beta^C +v_\gamma^C
\end{equation}
and 
\begin{equation}
\widetilde{H}_\alpha=H^0 + v_\alpha^s + v_\alpha^C + u_\alpha^C,
\label{htilde}
\end{equation}
where $u_\alpha^C$ is an auxiliary potential in 
coordinate $\eta_\alpha$,
which is required to have the asymptotic form 
\begin{equation}
u_\alpha^C \sim \frac{Z_\alpha (Z_\beta+Z_\gamma) }{\eta_\alpha}
\end{equation}
as ${\eta_\alpha \to \infty}$. In fact $u_\alpha^C$ 
is an effective Coulomb
interaction between the center of mass of the subsystem 
$\alpha$ (with
charge $Z_\beta+Z_\gamma$) and the third particle 
(with charge $Z_\alpha$).

Let us introduce the resolvent operators: 
\begin{equation}
G(z)=(z-H)^{-1},
\end{equation}
\begin{equation}
G_\alpha ^C(z)=(z-H_\alpha ^C)^{-1},
\end{equation}
\begin{equation}
\widetilde{G}_\alpha (z)=(z-\widetilde{H}_\alpha )^{-1}.
\end{equation}
The operator $G_\alpha ^C$ is Noble's channel 
Coulomb Green's operator
and $\widetilde{G}_\alpha $ is the channel 
distorted Coulomb Green's
operator introduced by Bencze \cite{benczenp}. 
These operators are connected via
the following resolvent relations: 
\begin{equation}
G(z)=G_\alpha ^C(z)+G_\alpha ^C(z)V^\alpha G(z),  \label{g3b}
\end{equation}
\begin{equation}
G_\alpha ^C(z)=\widetilde{G}_\alpha (z)+
\widetilde{G}_\alpha (z)U^\alpha
G_\alpha ^C(z),  \label{g2b}
\end{equation}
where $V^\alpha =v^s_\beta +v^s_\gamma $ and $U^\alpha
 =v_\beta ^C+v_\gamma
^C-u_\alpha ^C$.

In the potential $U^\alpha$ the Coulomb tail of 
$v_\beta^C + v_\gamma^C$ is
compensated by $u_\alpha^C$. As concerning the
 asymptotic motion $U^\alpha$ is
of short-range type, so the equation (\ref{g2b})
 is mathematically well-behaved. 
The scattering states 
\begin{equation}
| \Phi_\alpha^{C(\pm)} \rangle = 
\lim\limits_{\varepsilon\to 0} \mbox{i}%
\varepsilon G_\alpha^C (E\pm \mbox{i}\varepsilon) |
 \Phi_\alpha \rangle
\end{equation}
satisfy the Lippmann-Schwinger equations 
\begin{equation}
|\Phi_{\alpha}^{C(\pm)}\rangle = 
| \widetilde{\Phi}_{\alpha}^{(\pm)}\rangle
+ \widetilde{G}_\alpha(E \pm \mbox{i} 0) U_\alpha
|\Phi_{\alpha}^{C(\pm)}\rangle ,  \label{lsuwf}
\end{equation}
where 
\begin{equation}
| \widetilde{\Phi}_{\alpha}^{(\pm)}\rangle = 
\lim\limits_{\varepsilon\to 0} %
\mbox{i}\varepsilon \widetilde{G}_\alpha
 (E\pm \mbox{i}\varepsilon) |
\Phi_{\alpha} \rangle.  \label{phitild}
\end{equation}
In configuration-space representation the sates 
$| \widetilde{\Phi}%
_{\alpha}^{(\pm)}\rangle$ are given as 
\begin{equation}
\langle \xi_\alpha \eta_\alpha | 
\widetilde{\Phi}_{\alpha}^{(\pm)}\rangle =
\langle \eta_\alpha | 
\widetilde{\chi}_{\alpha}^{(\pm)}\rangle \langle
\xi_\alpha | \phi_{\alpha} \rangle,  \label{chipm}
\end{equation}
where $\langle \eta_\alpha | 
\widetilde{\chi}_{\alpha}^{(\pm)}\rangle$ are
scattering functions in the Coulomb-like potential $u_\alpha^C$.

In (\ref{g3b}) the potential $V^\alpha$ is of 
short-range type and $%
G_\alpha^C$ contains all the Coulomb interactions.
 Now, all the formulas
which exist in the conventional short-range three-body theory can
analogously be derived, only the channel Green's 
operator $G_\alpha$ has to be
replaced, \`{a} la Noble, by $G_\alpha^C$ throughout. 
One can analogously
perform the Faddeev decomposition and for 
the Faddeev components $|
\psi^{(\pm)} \rangle$ of the scattering function 
\begin{equation}
| \Psi_\alpha^{(\pm)} \rangle = 
\lim\limits_{\varepsilon\to 0} \mbox{i}%
\varepsilon G (E_\alpha \pm \mbox{i}\varepsilon)| 
\Phi_\alpha \rangle = | 
\Phi_\alpha \rangle + \sum\limits_{\gamma}|
\psi_\gamma^{(\pm)} \rangle 
\end{equation}
one arrives at the Faddeev-Noble integral equations \widetext
\begin{equation}
|\psi_{\alpha}^{(\pm)}\rangle=\delta_{\beta \alpha} |\Phi_{\alpha
m}^{C(\pm)}\rangle + G_\alpha^C (E \pm \mbox{i} 0) [ v^s_\alpha
|\psi_{\beta}^{(\pm)}\rangle + v^s_\alpha |
\psi_{\gamma}^{(\pm)}\rangle ]
\label{feqs}
\end{equation}
with a cyclic permutation in $\alpha, \beta, \gamma$.

The S-matrix elements of scattering processes can 
be obtained from the
resolvent of the total Hamiltonian by the reduction 
technique \cite{redu} 
\begin{equation}
S_{\beta n,\alpha m}=
\lim\limits_{t\to \infty }\lim\limits_{\varepsilon \to
0}\mbox{i}\varepsilon \mbox{e}^{\mbox{i}(E_{\beta n}-
E_{\alpha m})t}\langle
\Phi _{\beta n}|G(E_{\alpha m}+\mbox{i}\varepsilon )|
\Phi _{\alpha m}\rangle
.  \label{sm}
\end{equation}
The subscript $m$ and $n$ denote the $m$-th and 
$n$-th eigenstates of the
corresponding subsystems, respectively. 
If we substitute now (\ref{g3b})
into (\ref{sm}) we can get, like in \cite{benczenp}, 
the following two terms:
\begin{equation}
S_{\beta n,\alpha m} =S_{\beta n,\alpha m}^{(1,2)} +
S_{\beta n,\alpha m}^{(3)}= S_{\beta n,\alpha m}^{(1)} +
S_{\beta n,\alpha m}^{(2)} +S_{\beta n,\alpha m}^{(3)}
\end{equation}
\begin{equation}
S_{\beta n,\alpha m}^{(1,2)}=\lim\limits_{t\to \infty
}\lim\limits_{\varepsilon \to 0}\mbox{i}\varepsilon 
\mbox{e}^{\mbox{i}%
(E_{\beta n}-E_{\alpha m})t}\langle \Phi _{\beta n}|
G_\alpha ^C(E_{\alpha m}+%
\mbox{i}\varepsilon )|\Phi _{\alpha m}\rangle   \label{s12}
\end{equation}
\begin{equation}
S_{\beta n,\alpha m}^{(3)}=\lim\limits_{t\to \infty
}\lim\limits_{\varepsilon \to 0}\mbox{i}\varepsilon 
\mbox{e}^{\mbox{i}%
(E_{\beta n}-E_{\alpha m})t}\langle \Phi _{\beta n}|
G_\alpha ^C(E_{\alpha m}+%
\mbox{i}\varepsilon )V^\beta G(E_{\alpha m}+\mbox{i}
\varepsilon )|\Phi
_{\alpha m}\rangle .
\end{equation}
We substitute again (\ref{g2b}) into (\ref{s12}) and 
the first term
yields again two further terms 
\begin{equation}
S_{\beta n,\alpha m}^{(1)}=\lim\limits_{t\to \infty
}\lim\limits_{\varepsilon \to 0}\mbox{i}\varepsilon 
\mbox{e}^{\mbox{i}%
(E_{\beta n}-E_{\alpha m})t}\langle \Phi _{\beta n}|
\widetilde{G}_\alpha
(E_{\alpha m}+\mbox{i}\varepsilon )|\Phi _{\alpha m}\rangle 
\end{equation}
\begin{equation}
S_{\beta n,\alpha m}^{(2)}=\lim\limits_{t\to \infty
}\lim\limits_{\varepsilon \to 0}\mbox{i}\varepsilon 
\mbox{e}^{\mbox{i}%
(E_{\beta n}-E_{\alpha m})t}\langle \Phi _{\beta n}|
\widetilde{G}_\alpha
(E_{\alpha m}+\mbox{i}\varepsilon )U^\alpha 
G_\alpha ^C(E_{\alpha m}+\mbox{i}%
\varepsilon )|\Phi _{\alpha m}\rangle .
\end{equation}
\narrowtext
Making use of the properties of the resolvent operators 
the limits can be
performed and we arrive at the following, physically 
very plausible, result.
The first term, $S_{\beta n,\alpha m}^{(1)}$, is the 
S-matrix of a two-body
single channel scattering on the potential $u_\alpha ^C$ 
\begin{equation}
S_{\beta n,\alpha m}^{(1)}=\delta _{\beta \alpha }
\delta _{nm}S(u_\alpha ^C).
\label{s1}
\end{equation}
If $u_\alpha ^C$ is a pure Coulomb interaction
 $S(u_\alpha ^C)$ falls back
to the S-matrix of the Rutherford scattering. 
The second term, $S_{\beta
n,\alpha m}^{(2)}$, describes a two-body 
multichannel scattering on the
potential $U^\alpha $ 
\begin{equation}
S_{\beta n,\alpha m}^{(2)}=-2\pi \mbox{i}
\delta _{\beta \alpha }\delta
(E_{\beta n}-E_{\alpha m})\langle 
\widetilde{\Phi }_{\beta n}^{(-)}|U^\alpha
|\Phi _{\alpha m}^{C(+)}\rangle .  \label{s2}
\end{equation}
The third term contains of the complete three-body dynamics 
\begin{equation}
S_{\beta n,\alpha m}^{(3)}=-2\pi 
\mbox{i}\delta (E_{\beta n}-E_{\alpha
m})\langle \Phi _{\beta n}^{C(-)}|V^\beta |
\Psi _{\alpha m}^{(+)}\rangle .
\label{s3}
\end{equation}
Utilizing the properties of the Faddeev components 
\cite{gloecklebook}
the matrix elements in (\ref{s3}) can be rewritten 
in a form which is better
suited for numerical calculations 
\begin{equation}
\langle \Phi _{\beta n}^{C(-)}|V^\beta |
\Psi _{\alpha m}^{(+)}\rangle
=\sum_{\gamma \neq \beta }\langle 
\Phi _{\beta n}^{C(-)}|v^s_\beta |\psi
_{\gamma m}^{(+)}\rangle .  \label{s3v}
\end{equation}
We note, that if the Coulomb 
interactions are absent the whole
''three-potential'' formalism falls back to 
the conventional short-range
formalism.

\section{SOLUTION OF THE THREE-BODY INTEGRAL EQUATIONS}

To solve operator equations in quantum mechanics 
one needs a suitable
representation for the operators. For solving 
integral equations it is
especially advantageous if one uses such a 
representation where the Green's
operator is simple. The free Green's operator 
takes a very simple form in
momentum representation. This is the main 
reason why for the solution of
Faddeev equations, in the presence of short-range interactions,
momentum-representation techniques  perform  so 
successfully (see for a
recent review Ref.~\cite{gloeckle-report}). 
Since the momentum
representation is a continuous representation, 
to solve the equation one
needs also some kind of discretization.

For the two-body Coulomb Green's operator 
there exists a Hilbert-space basis
in which its representation is very simple, 
it is the Coulomb-Sturmian (CS)
basis. In this representation-space the 
Coulomb Green's operator can be
given by simple and well-computable analytic 
functions \cite{papp1}. This
basis is a countable set. If we represent the 
interaction term on a finite
subset of the basis it looks like a kind of 
separable expansion of the
potential, so the integral equation becomes 
an algebraic equation. The
completeness of the basis ensures the 
convergence of the method.

In the past few years along this idea we have 
developed a quantum-mechanical
approximation method for treating Coulomb-like 
interactions in two-body
calculations. Bound- and resonant-state calculations
 were presented first 
\cite{papp1}, then the method was extended to 
scattering states \cite{papp2}
and multichannel problems \cite{pzews}. Since only 
the asymptotically
irrelevant short range interaction is approximated, 
the correct (two-body)
Coulomb asymptotics is guaranteed. The corresponding
 computer codes for
solving two-body bound-, resonant- and scattering-state 
problems were also
published \cite{cpc}.

Recently the CS separable expansion approach was 
applied to solving the
three-body bound-state problem in the presence of 
short-range plus repulsive
Coulomb interactions \cite{pzwp}. The homogeneous 
Faddeev-Noble integral
equations were solved by expanding only the short-range 
part of the
interaction in a separable form while treating the 
long-range part in an
exact manner. The efficiency of the method was 
demonstrated in benchmark
calculations of the three-body bound-state problem
 without and with Coulomb
interactions. In both cases the solution showed a 
rapid convergence, and,
whenever a comparison were possible to existing 
results in the literature,
correct predictions for the binding energies and
 wave functions were
achieved. The method was also applied in realistic
 calculations \cite{leo}.

In subsection A below we will define the basis 
states in two- and
three-particle Hilbert space. In subsection B 
below we recapitulate some of
the most important formulas of the two-body 
problem (the details are given in
Refs.~\cite{papp1,papp2,cpc}), while in subsection C
below the solution of the three-body Coulomb 
scattering problem along the CS
separable expansion technique is presented.

\subsection{Basis states}

The CS functions, which are the solutions of 
the Sturm-Liouville problem of
hydrogenic systems \cite{rotenberg}, in some 
angular momentum state $l$ are
defined in configuration- and momentum-space as 
\begin{equation}
\langle r|nl\rangle =\left[ \frac{n!}{(n+2l+1)!}\right]
^{1/2}(2br)^{l+1}e^{-br}L_n^{2l+1}(2br)  \label{basisr}
\end{equation}
and 
\begin{eqnarray}
\langle p|nl\rangle 
 &=& \frac{2^{l+3/2}l!(n+l+1)\sqrt{n!}}{%
\sqrt{\pi (n+2l+1)!}}  \nonumber \\
&&\times \; \frac{b (2bp)^{l+1}}{(p^2+b^2)^{2l+2}}G_n^{l+1}
\left(\frac{p^2-b^2}{p^2+b^2}%
\right) ,  \label{basisp}
\end{eqnarray}
respectively, and $n=0,1,2,\ldots $. Here, $L$ and $G$ 
represent the
Laguerre and Gegenbauer polynomials, respectively, and 
$b$ relates to the
energy in the Sturm-Liouville equation. We take $b$ as 
a fixed real
parameter, thus working with energy-independent bound 
state CS functions.
In an angular momentum subspace they form a complete set 
\begin{equation}
{\bf {1}}=\lim\limits_{N\to \infty }\sum_{n=0}^N|
\widetilde{nl}\rangle
\langle nl|=\lim\limits_{N\to \infty }{\bf {1}}_N,  \label{unity}
\end{equation}
where $|\widetilde{nl}\rangle$ in configuration-space 
representation reads 
\begin{equation}
\langle r|\widetilde{nl}\rangle =\frac 1r\langle r|nl\rangle .
\end{equation}

The three-body Hilbert space is a direct sum of two-body 
Hilbert spaces.
Thus, the appropriate basis in angular momentum 
representation (omitting the
explicit spin and isospin dependence from our notation) 
should be defined as
a the direct product 
\begin{equation}
| n \nu l \lambda \rangle_\alpha = 
| n l \rangle_\alpha \otimes | \nu
\lambda \rangle_\alpha , \ \ \ \ (n,\nu=0,1,2,\ldots),  
\label{cs3}
\end{equation}
with the CS states from Eq.~(\ref{basisr}) or 
Eq.~(\ref{basisp}). Here $l$
and $\lambda$ denote the angular momenta of the 
two-body pair $(\beta,\gamma)
$ and of the third particle $\alpha$ relative to 
the center of mass of the
pair, respectively. Now the completeness relation 
takes the form (with
angular momentum summation implicitly included) 
\begin{equation}
{\bf 1} =\lim\limits_{N\to\infty} \sum_{n,\nu=0}^N |
 \widetilde{n \nu l
\lambda} \rangle_\alpha \ \mbox{}_\alpha\langle 
{n \nu l \lambda} | =
\lim\limits_{N\to\infty} {\bf 1}_{N}^\alpha
\end{equation}
where the configuration-space representation in 
terms of Jacobi coordinates $%
\xi_\alpha$ and $\eta_\alpha$ reads: 
\begin{equation}
\langle \xi_\alpha \eta_\alpha |
\widetilde{ n \nu l \lambda }
\rangle_\alpha= \frac{1}{\xi_\alpha \eta_\alpha} 
\langle \xi_\alpha
\eta_\alpha |{\ n \nu l \lambda }\rangle_\alpha .
\end{equation}
It should be noted that in the three-particle 
Hilbert space we can introduce
three equivalent bases which belong to fragmentation 
$\alpha$, $\beta$
and $\gamma$.

\subsection{Coulomb-Sturmian separable expansion in 
two-body problems}

Let us study a two-potential case of short-range
 plus Coulomb-like
interactions 
\begin{equation}
v_l=v^s_l+v^C
\end{equation}
and consider the inhomogeneous Lippmann-Schwinger 
equation for the
scattering state $|\psi _l\rangle $ in some partial wave $l$ 
\begin{equation}
|\psi _l\rangle =|\phi _l^C\rangle +g_l^C(E)v^s_l|\psi _l\rangle .  
\label{LS}
\end{equation}
Here $|\phi _l^C\rangle $ is the regular Coulomb function, 
$g_l^C(E)$ is the
two-body Coulomb Green's operator 
\begin{equation}
g_l^C(E)=(E-h_l^0-v^C)^{-1}
\end{equation}
with the free Hamiltonian denoted by $h_l^0$. We make
 the following
approximation on Eq.~(\ref{LS}) 
\begin{equation}
|\psi _l\rangle =|\varphi _l^C\rangle +
g_l^C(E){\bf {1}}_N v^s_l {\bf {1}}%
_N|\psi _l\rangle ,  \label{LSapp}
\end{equation}
i.e.\ we approximate the short-range potential 
$v_l^s $ by a separable form 
\begin{equation}
v_l^s\approx \sum_{n, n^{\prime } =0}^N
|\widetilde{
n l}\rangle  \;
\underline{v}_{l}^s\;\mbox{}%
\langle \widetilde{n^{\prime }
l }|  \label{sepfe2b}
\end{equation}
where 
\begin{equation}
\underline{v}_{l}^s=
\langle n l|
v_l^s| n^{\prime } l \rangle .  
\label{v2b}
\end{equation}
Multiplied with the CS states $\langle \widetilde{nl}|$
 from the left, Eq.~(\ref
{LSapp}) turns into a linear system of equations 
for the wave-function
coefficients $\underline{\psi }_{ln}=\langle 
\widetilde{nl}|\psi _l\rangle $%
\begin{equation}
\lbrack (\underline{g}_l^C(E))^{-1}-
\underline{v}_l^s]\underline{\psi }_l=%
\underline{\varphi }_l^C,  \label{eq18a}
\end{equation}
where 
\begin{equation}
\underline{\varphi }_{ln}^C=\langle 
\widetilde{nl}|\varphi _l^C\rangle 
\label{phiov}
\end{equation}
and
\begin{equation}
\underline{g}_{lnn^{\prime }}^C(E)=
\langle \widetilde{nl}|g_l^C(E)|%
\widetilde{n^{\prime }l}\rangle .  \label{gcme}
\end{equation}
While the matrix elements of the potential may 
be evaluated (numerically) for any
given short-range potential either in 
configuration or in momentum space,
the matrix elements (\ref{gcme}) and 
the overlap (\ref{phiov}) can be
calculated analytically \cite{papp1}; 
the corresponding computer code is
available from Ref.~\cite{cpc}. This 
fact then also allows to calculate the
matrix elements of the full Green's 
operator in the whole complex plane, 
\begin{equation}
\underline{g}_l(z)=((\underline{g}_l^C(z))^{-1}-
\underline{v}_l^s)^{-1},
\label{2bgreen}
\end{equation}
this will be needed later on in the solution 
of the three-body problem with
charged particles. Of course, bound-state 
solutions can also be generated by
solving the homogeneous version of Eq.~(\ref{eq18a}).

\subsection{Coulomb-Sturmian separable expansion 
approach to three-body
Coulomb scattering problems}

In the set of Faddeev-Noble equations (\ref{feqs}) 
we make the following
approximation: 
\begin{equation}
|\psi _\alpha \rangle =\delta _{\beta \alpha }|
\Phi _{\alpha m}^C\rangle
+G_\alpha ^C[{\bf 1}_N^\alpha v^s_\alpha 
{\bf 1}_N^\beta |\psi _\beta \rangle +
{\bf 1}_N^\alpha v^s_\alpha {\bf 1}_N^\gamma 
|\psi _\gamma \rangle ],
\label{feqsapp}
\end{equation}
\narrowtext \noindent
i.e.\ we approximate the short-range potential 
$v_\alpha^s $ in the three-body
Hilbert space by a separable form 
\begin{equation}
v_\alpha ^s\approx \sum_{n,\nu ,n^{\prime },
\nu ^{\prime }=0}^N|\widetilde{%
n\nu l\lambda }\rangle _\alpha \;
\underline{v}_{\alpha \beta }^s\;\mbox{}%
_\beta \langle \widetilde{n^{\prime }
\nu ^{\prime }l^{\prime }\lambda
^{\prime }}|  \label{sepfe}
\end{equation}
where 
\begin{equation}
\underline{v}_{l_\alpha \lambda _\alpha n\nu ,
{l^{\prime }}_\beta {\lambda
^{\prime }}_\beta n^{\prime }\nu ^{\prime }}^s=
(1-\delta _{\alpha \beta })\ %
\mbox{}_\alpha \langle n\nu l\lambda |
v_\alpha^s|n^{\prime }\nu ^{\prime
}l^{\prime }{\lambda }^{\prime }\rangle_\beta .  
\label{vab}
\end{equation}
In (\ref{sepfe}) the ket and bra states are belonging to different 
fragmentations depending on the
environments of the potential operators in the equations.

Muliplied with the CS states 
$\mbox{}_\alpha \langle \widetilde{n\nu l\lambda }|
$ from the left, Eqs. (\ref{feqsapp}) 
turn into a linear system of equations
for the coefficients of the Faddeev 
components $\underline{\psi }_{l_\alpha
\lambda _\alpha n\nu }=\mbox{}_\alpha \langle 
\widetilde{n\nu l\lambda }%
|\psi _\alpha \rangle $: 
\begin{equation}
\lbrack (\underline{G}^C)^{-1}-
\underline{v}^s]\underline{\psi }=\underline{%
\Phi }^C,  \label{fep1}
\end{equation}
with 
\begin{equation}
\underline{G}_{l_\alpha \lambda _\alpha n\nu ,
{l^{\prime }}_\alpha {\lambda
^{\prime }}_\alpha n^{\prime }\nu ^{\prime }}^C=
\delta _{\alpha \beta }\ %
\mbox{}_\alpha \langle \widetilde{n\nu l\lambda }|
G_\alpha ^C|\widetilde{%
n^{\prime }\nu ^{\prime }{l^{\prime }}
{\lambda ^{\prime }}}\rangle _\alpha ,
\label{G}
\end{equation}
and 
\begin{equation}
\underline{\Phi }_{l_\alpha \lambda _\alpha n\nu }^C=
\mbox{}_\alpha \langle 
\widetilde{n\nu l\lambda }|\Phi _\alpha ^C\rangle . 
 \label{P}
\end{equation}
Notice that the matrix elements of the Green's 
operator are needed only
between the same partition $\alpha $ whereas 
the matrix elements of the
potentials occur only between different 
partitions $\alpha $ and $\beta $.
The latter may again be evaluated numerically 
either in configuration or
momentum space by making use of the transformation 
of Jacobi coordinates 
\cite{bb}.

Unfortunately neither the matrix elements (\ref{G}) 
nor the overlap (\ref{P}%
) are known. However, Eqs.\ (\ref{g2b}) and 
(\ref{lsuwf}), which are, in
fact, two-body Lippmann-Schwinger equations, 
link them to relatively simpler
quantities. If we perform again the separable 
approximation on potential $%
U^\alpha$ with the help of the formal solution 
of (\ref{g2b}) we may now
express the inverse matrix 
$(\underline{G}^C_\alpha (E))^{-1}$ as 
\begin{equation}
(\underline{G}^C_\alpha )^{-1}= 
(\underline{\widetilde{G}}_\alpha )^{-1} -%
\underline{U}^\alpha,
\end{equation}
where 
\begin{equation}
\underline{\widetilde{G}}_{l_\alpha \lambda_\alpha n \nu , 
l^{\prime}_\alpha 
{\lambda}^{\prime}_\alpha n^{\prime}\nu^{\prime}} =
 \mbox{}_\alpha\langle n
\nu l \lambda | \widetilde{G}_\alpha |
 n^{\prime}\nu^{\prime}l^{\prime}{%
\lambda}^{\prime}\rangle_\alpha  \label{gtilde}
\end{equation}
and 
\begin{equation}
\underline{U}^\alpha_{l_\alpha \lambda_\alpha n \nu ,
 l^{\prime}_\alpha {%
\lambda}^{\prime}_\alpha n^{\prime}\nu^{\prime}} =
 \mbox{}_\alpha\langle n
\nu l \lambda | U^\alpha | n^{\prime}\nu^{\prime}
l^{\prime}{\lambda}%
^{\prime}\rangle_\alpha.
\end{equation}
In a similar way, with the help of the formal
 solution of (\ref{lsuwf}) we
get 
\begin{equation}
\underline{\Phi}_\alpha^C = [
 (\underline{\widetilde{G}}_\alpha )^{-1} -%
\underline{U}^\alpha ]^{-1} 
(\underline{\widetilde{G}}_\alpha )^{-1} 
\underline{\widetilde{\Phi}}_\alpha,
\end{equation}
where 
\begin{equation}
\underline{\widetilde{\Phi}}_\alpha= 
\mbox{}_\alpha\langle n \nu l \lambda | 
\widetilde{\Phi}_\alpha \rangle.
\end{equation}

The state $| \widetilde{\Phi}_\alpha \rangle$, 
in fact, is a product of a
two-body bound-state wave function in coordinate 
$\xi_\alpha$ and a two-body
scattering-state wave function in coordinate 
$\eta_\alpha$. Their CS
representations are known from the 
two-particle case of the previous section
[cf.\ Eq. (\ref{eq18a})].

\begin{figure}[tbp]
\psfig{file=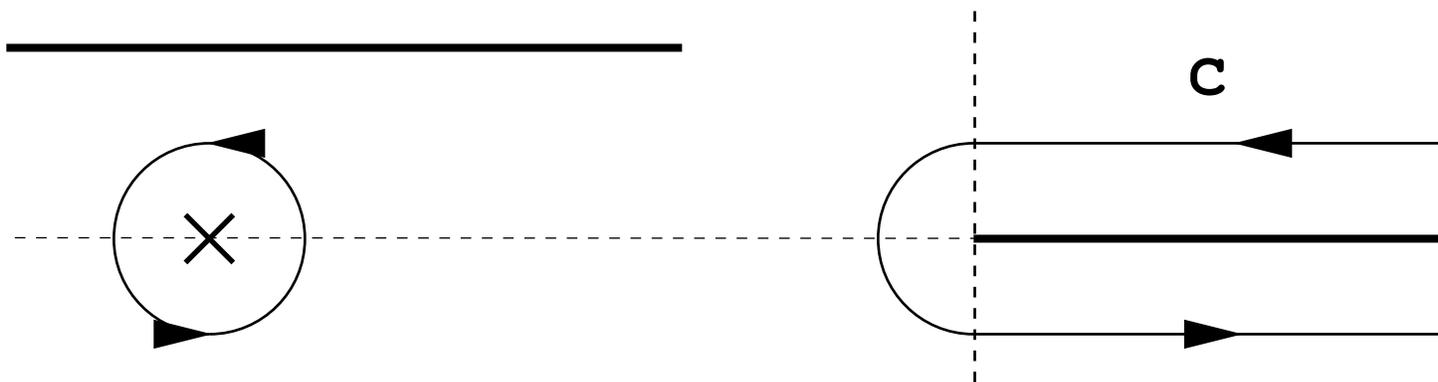}
\caption{ Contour $C$ for the integral for 
$\widetilde{G}_\alpha (E+\mbox{i}
\varepsilon)$ in case of the three-body scattering problem. 
The contour $C$
encircles the continuous and discrete spectrum of 
$h_{\xi_\alpha}$. In the $%
\varepsilon \to 0$ limit the topology 
of the contour should be kept.}
\label{fig}
\end{figure}

For the calculation of the matrix 
elements in Eq.\ (\ref{gtilde}) we proceed
in a similar way as in the case of
 three-body bound-states \cite{pzwp}. Since in $%
\widetilde{H}_\alpha $ of Eq.\ (\ref{htilde}) 
we can write the
three-particle free Hamiltonians as a sum of two-particle
free Hamiltonians 
\begin{equation}
H^0=h_{\xi _\alpha }^0+h_{\eta _\alpha }^0,
\end{equation}
the Hamiltonian $\widetilde{H}%
_\alpha $ appears as a sum of two Hamiltonians 
acting on different coordinates 
\begin{equation}
\widetilde{H}_\alpha =h_{\xi _\alpha }+h_{\eta _\alpha },
\end{equation}
with $h_{\xi _\alpha }=
h_{\xi _\alpha }^0+v_\alpha ^s(\xi _\alpha )+v_\alpha
^C(\xi _\alpha )$ and $h_{\eta _\alpha }=
h_{\eta _\alpha }^0+u_\alpha
^C(\eta _\alpha )$, which, of course, commute. 
Thus we can apply the
convolution theorem \cite{bianchi} 
\begin{eqnarray}
\widetilde{G}_\alpha (z)=(z-h_{\xi _\alpha }-
h_{\eta _\alpha })^{-1} &=&%
\frac 1{2\pi i}\oint_C dw(z-w-h_{\eta _\alpha })^{-1}  
\nonumber \\
&&\times (w-h_{\xi _\alpha })^{-1}. 
 \label{contourint}
\end{eqnarray}
Here the contour $C$ should encircle, in positive direction, the 
spectrum of $h_{\xi _\alpha }$
without penetrating into the spectrum of $h_{\eta _\alpha }$ . For
scattering-state energies at real $z$
 these singularities overlap. To find
the correct path one should take the $z=E+\mbox{i}\varepsilon $ 
case with
finite $\varepsilon $. Now the condition on $C$ 
can easily be fulfilled and then,
one should take the $\varepsilon \to 0$ 
limit allowing only analytic
deformation for the contour $C$ [see Fig. \ref{fig}].

After sandwiching the above Green's operator between the 
CS states, the
integral in Eq. (\ref{contourint}) appears in the form 
\begin{eqnarray}
\lefteqn{\underline{\widetilde{G}}_{l_\alpha 
\lambda_\alpha n \nu ,
l_\alpha' \lambda_\alpha' n' \nu' } (E+\mbox{i} 0) } 
 \nonumber \\
& =&\frac{1}{2 \pi i} \oint_C dw \ \mbox{}_\alpha\langle 
\widetilde{ \nu \lambda }|
(E+\mbox{i} 0 -w - h_{\eta_\alpha})^{-1} | \widetilde{
\nu^{\prime}{\lambda^{\prime}} }
\rangle_\alpha \nonumber \\
&& \ \times \ \mbox{}_\alpha\langle
 \widetilde{ n l}| (w -
h_{\xi_\alpha})^{-1} |
\widetilde{ n^{\prime}{l^{\prime}}}
\rangle_\alpha,  \label{contourint2}
\end{eqnarray}
where both matrix elements occurring in 
the integrand are known from
the two-particle case [cf. Eq. (\ref{2bgreen})].

\section{TESTS OF THE METHOD}

In this section we demonstrate the performance 
of the method in calculations
of three-body short-range and Coulombic scattering 
phase shifts at energies
below the breakup threshold. We have selected cases 
that serve as benchmarks
for various three-body scattering calculations. 
As an example we take a
model three-nucleon problem with s-wave Malflet-Tjon (MT) I-III potential, 
acting in singlet
and triplet states, as parametrized in Ref.~\cite{chen}. 
We have calculated
quartet and doublet $n-d$ and $p-d$ phase shifts and 
compare them to the
results of the configuration-space Faddeev 
calculations of Ref.~\cite{chen}.

Before presenting the final results, let us 
demonstrate the convergence of
the results for scattering phase shifts at 
various energies. We take extreme
cases, one is at very low energy, an other one is 
jut below the
breakup threshold, the third one is in between.
We select two-channel doublet 
$n-d$ and $p-d$ cases,
because this case is more complicated than 
the one-channel quartet case.
Tables \ref{tabconv1} and \ref{tabconv2} show 
that convergence up 
to 4 significant digits can
comfortably achieved with $N=30$ terms applied
 for $n$ and $\nu $ in the
separable expansion. Remarkably, the speed of 
convergence is everywhere
similar, irrespective of energy and whether or
not Coulomb forces are present.

In Tables \ref{tabnd} and \ref{tabpd} we compare 
our converged results to
the configuration-space Faddeev calculations of 
the Los Alamos-Iowa group 
\cite{chen}. We can report perfect agreements 
in all cases.

In  Tables \ref{tabconv2} and \ref{tabpd} and also 
in Ref.~\cite{chen}
the Coulomb modified 
phase shift from short-range plus polarization
potential ($\delta^{c,ps}$) were presented, and the
corresponding values were used in the calculation of
the scattering lengths $a_{pd}^{c,ps}$.
However, since our three-potential
formalism allows a unique separation of these two effects
we can also calculate the Coulomb plus polarization modified
short-range phase shift $\delta^{cp,s}$
and the corresponding scattering lengths $a_{pd}^{cp,s}$.
Theoretically the scattering length should
be calculated from $\delta^{cp,s}$ since $a_{pd}^{c,ps}$
is minus infinite \cite{cps}. In practical calculations
the extrapolation to zero energy were made from
higher energy and this minus infinity limit were not seen.
Careful analyses of low-energy $p-d$ calculations indicated
that in these condition $a_{pd}^{c,ps}$ is a good
approximation to $a_{pd}^{cp,s}$ \cite{chen,bzank,cps,matel}.
We have calculated the scattering length from 
both phase shifts using the
formula of Ref.~\cite{cps}
\begin{equation}
a=\lim\limits_{k\to 0} a(k) =
\lim\limits_{k\to 0} -\frac{\tan \delta(k)}{k C^2(k)},
\end{equation}
where $k$ is the wavenumber and 
\begin{equation}
C^2(k)=\frac{2 \pi \eta}{\mbox{e}^{2 \pi \eta}-1}
\end{equation}
with $\eta=m e^2/\hbar k$, $m$ beeing the reduced mass.
From $a(k)$ values correspondig to energies
down to 0.01 MeV in the doublet case and to 0.001 MeV
in the quartet case, where the phase shift
still can reliably be calculated, 
we extrapolated to zero energy.
The results are given in Table \ref{tabcps}.
We can see that the effect of the polarization potential
can realy be neglected in the calculation of the 
$p-d$ scattering lengths.

Besides the number of terms in the expansion there 
is only one parameter in
the method, the $b$ parameter of the basis. 
This should be chosen according
to the range of the potential. We have found 
that the converged results do
not depend on $b$ and for a wide range of 
reasonable values even the speed of the
convergence is rather insensitive to the choice of $b$.
In all calculations presented the same value was applied 
($b=1 \mbox{fm}^{-1}$).

\section{CONCLUSION}

We have suggested a three-potential formalism for 
treating the three-body
Coulomb scattering problem. In absence of Coulomb
 interactions the formalism
falls back to the usual short-range formalism.
 According to the
three-potential picture the three-body Coulomb
 scattering starts with a
two-body single channel Coulomb scattering, then 
it goes over to a two-body
multichannel scattering on the intermediate-range 
polarization potential,
finally comes the three-body scattering due to the
 short-range potentials.
This formalism preserves the mathematical correctness 
of   Noble's
approach, and along the idea of channel distorted 
formalism of Bencze,
without neglecting important terms, gives solvable
 equations.

These ''solvable'' equations are certainly too 
complicated for most of the
numerical methods available in the literature, 
but the Coulomb-Sturmian
separable expansion method can cope with them. 
It solves the three-body
integral equations by expanding only the 
short-range part of the interaction
in a separable form on a Hilbert-space basis 
while treating the long-range
part in an exact manner via a proper integral 
representation of the
three-body channel distorted Coulomb Green's operator.
 As a consequence the
method has good convergence properties and can in 
practice be made
arbitrarily accurate by employing an increasing 
number of terms in the
expansion. The usage of the Coulomb-Sturmian basis 
is essential as it allows
an exact analytic representation of the two-body 
Green's operator, and thus
the contour integral for the channel distorted 
Coulomb Green's operator can
be calculated also in the practice.

We have presented below-breakup calculations and 
got perfect agreements with
existing benchmark results. We have observed a fast
 convergence with respect
to increasing the number of terms in the expansion.
 Using high rank
expansion we got very accurate results. The 
quality of the convergence is
practically the same in short-range and in 
Coulomb case, Coulombic
calculations need only roughly $30\%$ more 
computer time. Although the
example presented here is rather simple, 
but not unrealistic, the method can
handle more complicated potentials, as were 
demonstrated in bound state
calculations \cite{pzwp,leo}.

Certainly, the toughest problems in nonrelativistic 
three-body scattering are the
above-breakup calculations with Coulomb interactions. 
In this respect, the
method presented here, is very promising since 
the equations used are
mathematically well-behaved also for this case.
 The extension of the contour
integral for above-breakup energies is straightforward.
 The only foreseeable
problem is that the interaction volume is much 
bigger and one needs much
higher terms in the expansion. Indeed, test
 calculations show that for
energies just a little bit above the breakup 
threshold with terms up to $N=34
$ we can reach acceptable convergence and good 
agreements with benchmark
results \cite{bench}, but the method fails to
 reach convergence for higher
energies. This indicates that the mathematical 
formulation is correct, but
the available computing power is not sufficient. 
So, the method needs
some more polishing and we have to think a little bit further.

\acknowledgments

This work has been supported by OTKA under
contract T17298.

\newpage

\begin{table}[tbp]
\caption{Convergence of the $\mbox{}^2\delta_{nd}$ 
phase shifts 
for three-nucleon system interacting 
via the MT I-III potential at various energies,
with increasing basis for the separable expansion. 
$N$ denotes the maximum
number of basis states employed for $n$ and $\nu$. 
The phase shifts are in degrees.}
\label{tabconv1}
\begin{tabular}{rccc}
& \multicolumn{3}{c}{$\mbox{}^2\delta_{nd}$}  \\ 
$N$ & 0.1 MeV & 1.0 MeV & 2.18 MeV  \\ \hline
10 & -4.0908 & -20.704 & -32.896  \\ 
11 & -3.6803 & -20.330 & -33.596  \\ 
12 & -3.8876 & -20.450 & -33.599  \\ 
13 & -3.4583 & -20.630 & -33.901  \\ 
14 & -3.5189 & -20.562 & -33.963  \\ 
15 & -3.3917 & -20.629 & -33.834  \\ 
16 & -3.3434 & -20.653 & -33.723  \\ 
17 & -3.3293 & -20.636 & -33.594  \\ 
18 & -3.2914 & -20.660 & -33.525  \\ 
19 & -3.2940 & -20.652 & -33.341  \\ 
20 & -3.2764 & -20.654 & -33.282  \\ 
21 & -3.2783 & -20.656 & -33.267  \\ 
22 & -3.2709 & -20.653 & -33.290  \\ 
23 & -3.2715 & -20.655 & -33.326  \\ 
24 & -3.2688 & -20.654 & -33.377  \\ 
25 & -3.2688 & -20.654 & -33.424  \\ 
26 & -3.2681 & -20.654 & -33.460  \\ 
27 & -3.2678 & -20.654 & -33.483  \\ 
28 & -3.2678 & -20.654 & -33.491  \\ 
29 & -3.2676 & -20.654 & -33.486  \\ 
30 & -3.2676 & -20.654 & -33.473  \\ 
31 & -3.2675 & -20.654 & -33.456  \\ 
32 & -3.2675 & -20.654 & -33.438  \\ 
33 & -3.2675 & -20.654 & -33.423  \\ 
34 & -3.2675 & -20.654 & -33.413 
\end{tabular}
\end{table}

\begin{table}[tbp]
\caption{Convergence of the $\mbox{}^2\delta_{pd}$
phase shifts 
for three-nucleon system interacting 
via the MT I-III potential at various energies,
with increasing basis for the separable expansion. 
$N$ denotes the maximum
number of basis states employed for $n$ and $\nu$. 
The phase shifts are in
degrees.}
\label{tabconv2}
\begin{tabular}{rccc}
& \multicolumn{3}{c}{$\mbox{}^2\delta_{pd}$} \\ 
$N$ &  0.1 MeV & 1.0 MeV & 2.0 MeV \\ \hline
10 &  -0.9485 & -16.376 & -28.405 \\ 
11 &  -0.7947 & -15.924 & -28.659 \\ 
12 &  -0.8794 & -16.089 & -28.564 \\ 
13 &  -0.6721 & -16.195 & -28.847 \\ 
14 &  -0.7015 & -16.138 & -28.874 \\ 
15 &  -0.6241 & -16.210 & -28.840 \\ 
16 &  -0.5962 & -16.226 & -28.865 \\ 
17 &  -0.5817 & -16.216 & -28.840 \\ 
18 &  -0.5575 & -16.241 & -28.802 \\ 
19 &  -0.5572 & -16.232 & -28.800 \\ 
20 &  -0.5446 & -16.238 & -28.774 \\ 
21 &  -0.5455 & -16.238 & -28.767 \\ 
22 &  -0.5390 & -16.236 & -28.764 \\ 
23 &  -0.5394 & -16.238 & -28.761 \\ 
24 &  -0.5362 & -16.236 & -28.765 \\ 
25 &  -0.5363 & -16.237 & -28.768 \\ 
26 &  -0.5350 & -16.237 & -28.772 \\ 
27 &  -0.5349 & -16.237 & -28.776 \\ 
28 &  -0.5345 & -16.237 & -28.778 \\ 
29 &  -0.5343 & -16.237 & -28.780 \\ 
30 &  -0.5342 & -16.237 & -28.780 \\ 
31 &  -0.5341 & -16.237 & -28.780 \\ 
32 &  -0.5341 & -16.237 & -28.779 \\ 
33 &  -0.5340 & -16.237 & -28.778 \\ 
34 &  -0.5340 & -16.237 & -28.778
\end{tabular}
\end{table}

\begin{table}[tbp]
\caption{$\mbox{}^4\delta_{nd}$ and 
$\mbox{}^2\delta_{nd}$ phase shifts for
three-nucleon system interacting via 
the MT I-III potential at various
energies. The phase shifts are in degrees.}
\label{tabnd}
\begin{tabular}{lcccc}
& \multicolumn{2}{c}{$\mbox{}^4\delta_{nd}$} &
 \multicolumn{2}{c}{$\mbox{}%
^2\delta_{nd}$} \\ 
$E$ & Ref.~\cite{chen} & This work & 
Ref.~\cite{chen} & This work \\ \hline
0.001 & -2.09 & -2.092 & -0.230 & -0.229 \\ 
0.05 & -14.6 & -14.62 & -1.99 & -1.988 \\ 
0.1 & -20.4 & -20.44 & -3.28 & -3.267 \\ 
0.2 & -28.3 & -28.29 & -5.68 & -5.670 \\ 
0.3 & -34.0 & -33.98 & -7.95 & -7.944 \\ 
0.4 & -38.5 & -38.54 & -10.1 & -10.09 \\ 
0.5 & -42.4 & -42.37 & -12.1 & -12.12 \\ 
0.6 & -45.7 & -45.69 & -14.0 & -14.03 \\ 
0.7 & -48.6 & -48.63 & -15.8 & -15.83 \\ 
0.8 & -51.2 & -51.27 & -17.5 & -17.53 \\ 
0.9 & -53.6 & -53.66 & -19.1 & -19.13 \\ 
1.0 & -55.8 & -55.86 & -20.7 & -20.65 \\ 
1.633 & -66.7 & -66.72 & -28.6 & -28.60 \\ 
2.180 & -73.6 & -73.6 & -33.6 & -33.4
\end{tabular}
\end{table}

\begin{table}[tbp]
\caption{$\mbox{}^4\delta_{pd}$ and 
$\mbox{}^2\delta_{pd}$ phase shifts for
three-nucleon system interacting via 
the MT I-III potential at various
energies. The phase shifts are in degrees.}
\label{tabpd}
\begin{tabular}{lcccc}
& \multicolumn{2}{c}{$\mbox{}^4\delta_{pd}$} & 
\multicolumn{2}{c}{$\mbox{}%
^2\delta_{pd}$} \\ 
$E$ & Ref.~\cite{chen} & This work & 
Ref.~\cite{chen} & This work \\ \hline
0.001 & 0.0 & 0.0 & 0.0 & 0.0 \\ 
0.05 & -2.69 & -2.694 & -0.113 & -0.112 \\ 
0.1 & -7.46 & -7.458 & -0.537 & -0.534 \\ 
0.2 & -15.6 & -15.56 & -1.96 & -1.949 \\ 
0.3 & -21.9 & -21.86 & -3.73 & -3.720 \\ 
0.4 & -27.0 & -27.00 & -5.62 & -5.612 \\ 
0.5 & -31.3 & -31.34 & -7.53 & -7.520 \\ 
0.6 & -35.1 & -35.11 & -9.40 & -9.394 \\ 
0.7 & -38.4 & -38.43 & -11.2 & -11.21 \\ 
0.8 & -41.4 & -41.40 & -13.0 & -12.96 \\ 
0.9 & -44.1 & -44.09 & -14.6 & -14.63 \\ 
1.0 & -46.5 & -46.55 & -16.2 & -16.24 \\ 
0.667 & -37.3 & -37.37 & -10.6 & -10.62 \\ 
1.333 & -53.5 & -53.49 & -21.1 & -21.08 \\ 
2.0 & -63.8 & -63.74 & -28.8 & -28.78
\end{tabular}
\end{table}

\begin{table}[tbp]
\caption{$p-d$ scattering lengths for
three-nucleon system interacting via 
the MT I-III potential.}
\label{tabcps}
\begin{tabular}{ccc}
  &  This work & Ref.~\cite{chen} \\ \hline
$\mbox{}^4 \delta_{pd}^{c,ps}$ & 13.76 & 13.8  \\
$\mbox{}^4 \delta_{pd}^{cp,s}$ & 13.79 &        \\
$\mbox{}^2 \delta_{pd}^{c,ps}$ & 0.161 & 0.17  \\
$\mbox{}^2 \delta_{pd}^{cp,s}$ & 0.195 &        
\end{tabular}
\end{table}

\end{document}